\documentclass[runningheads,a4paper]{llncs}

\usepackage{amssymb}
\usepackage{graphicx}

\usepackage[ruled]{algorithm2e}
\usepackage{amsmath}

\SetAlFnt{\small}
\SetAlCapFnt{\small}
\SetAlCapNameFnt{\small}
\SetAlCapHSkip{0pt}
\IncMargin{-\parindent}

\usepackage[numbers]{natbib}

\begin{document}

\newcommand{\RR}{\mathbb{R}}
\newcommand{\EE}{\mathbb{E}}
\newcommand{\inner}[2]{\left\langle #1, #2 \right\rangle }

\newcommand{\bq}{\bar{q}}
\newcommand{\bp}{\bar{p}}
\newcommand{\Acal}{\mathcal{A}}
\newcommand{\Ccal}{\mathcal{C}}
\newcommand{\Rcal}{\mathcal{R}}
\newcommand{\ones}{\mathbf{1}}

\newcommand{\ie}{\emph{i.e.}}
\newcommand{\TODO}[1]{{\color{red} TODO: #1}}


\title{Bandit Market Makers}
\author{Nicol\'as Della Penna
\institute{ANU, NICTA}
\and Mark D. Reid
}

\maketitle

\begin{abstract}

We introduce a modular framework for market making. It combines cost-function based automated market makers with bandit algorithms. We obtain worst-case profits guarantee's relative to the best in hindsight within a class of natural "overround"  cost functions  . This combination allow us to have distribution-free guarantees on the regret of profits while preserving the bounded worst-case losses and computational tractability over combinatorial spaces of the cost function based approach. We present simulation results to better understand the practical behaviour of market makers from the framework.

\end{abstract}





\section{Introduction}

We propose a framework for profit-driven market making on state-contingent claims. It provides distribution-free guarantees on the regret of worst-case profits within those that can be obtained by a natural  class of "overround" cost function-based automated market makers. Worst-case profits are natural for a distribution-free analysis over outcomes, while "overround" cost functions are natural way to extend cost-functions to be able to obtain profits . The "overround" can be considered as the price the market maker charges for providing a unit of liquidity, since it is the cost of purchasing continent claims to obtain a unit payoff under any state of the world. 

The market makers pricing problem for assets with state-contingent payoffs can be decomposed into two sub-problems;

\begin{enumerate}
\item increasing the price of an asset as the proportion of outstanding contracts in it increases relative to those of other assets, so as to limit potential losses from it's outcome occurring. 
\item Finding the magnitude of the sum of prices that maximises the profits that can be gained by the market maker, the overround.
\end{enumerate}
 
The first sub-problem is well handled by the cost function literature, providing strong worst-case guarantees. We provide an overview of the mechanics of the cost function approach in \ref{sec:cost-functions}.  The second sub-problem has a explore vs exploit tradeoff at it's core: the  higher the prices for assets caused by a larger overround while increasing the profitability of a given set of contracts sold, reduces the demand for contracts. Conversely, lower prices due to a smaller overround reduce the profitability of any given contract bundle sold but can lead to higher demand for contracts. There is thus no a priori correct size for the overround. Given a choice of overround the market maker only observes the demand for assets at that level, so there is a need for it to explore the profitability of various levels that the overround can be set at, while selecting the profit maximising overround as it becomes apparent with sufficiently high probability. This exploration vs exploration tradeoff is well handled by the literature on bandit problems.

We consider a setting where on each unit of time the market maker commits to using a given overround cost function, and changes this cost function between time periods. The framework works by mapping the size of the overround into the bandit's arm, and the change in worst case profits in a given period to the rewards of the bandit algorithm. This combination allow us to have distribution-free guarantees from the bandit literature and computational tractability over combinatorial spaces of the cost-function based approach.  The modular approach has the advantage that it allows us to build on results in these two growing literatures, and to bring future developments in their algorithms and analisys into to market making domain.

\subsection{Examples}

To motivate and clarify the two aspects,  price discovery with bounded loss, and profit-maximising sum of assetsprices (overround) it is useful to consider two extreme situations the market maker can be faced with.

To understand the importance of the cost cost function, one can consider a situation in which in each period there is a trader who has perfect knowledge of the state of the world that will occur. It is important that even as in each time period trader buys the asset that will occur, in any quantity, the total exposure of the market maker be bounded, so as to be able to bound it's worst case loss. In this situation the optimal market maker is one that charges a price of 1 for all assets, since he incurs 0 loss, and there is no way to profit from the sequence of identical traders with perfect knowledge of the outcome. 

To understand the importance of the explore/exploit tradeoff in finding the profit maximising sum of prices, one can consider the situation on which traders are drawn form a symmetric belief distribution with a known centre and the relevant state of the world is a binary variable. The market maker must still learn the right amount to separate the price of the two assets so as to extract maximal profits. If we further specialise the situation to one where the beliefs are fully concentrated on two points.  At each time period we have a choice ov how far from the known centre to separate the two prices (this is equivalent to the overround), and we either observe the full demand, and thus know that it may still be possible to have a higher overround while attracting the same amount of trade, or we observe no trade and thus learn that the optimal overround is smaller.

\section{Related Literature}\label{sec:lit}

The market maker as a way to elicit information from traders has been studied in an extensive literature on cost function based automated market makers motivated
by prediction markets \cite{Hanson:2003,Hanson:2007,Chen:2008}.  By equating outcomes in the market setting with experts in the learning setting, and the
trades made in the market with the losses observed by the learning algorithm \citet{Chen:2010, abernethy2012efficient} a strikeing mathematical equivalence exists between cost function based prediction markets and regularised follow the leader online learning. Further work has shown has models in models with traders that have stochastically drawn beliefs correspond to mirror descent  \cite{Frongillo:2012}. The learning that takes place in these cases can be seen to be taking place over the probabilities of the events. In these cases the market maker can be seen as subsidising trade in securities contingent on an outcome the traders might not have exogenous reasons to hedge onto, as a way of paying to extract their beliefs about the likelihood of the events.  

The market maker as an agent motivated to make profits by offering liquidity to traders and attempting to maximise profit also has an extensive literature, through it has until recently remained separate from that on prediction market making. In \cite{das2008adapting} an optimal market maker when trader beliefs are symmetric is presented, in \cite{chakraborty2011market} the price process is assumed to be mean reverting. Learning again can be considered to be taking place in this case, but the learning is over buying and selling prices will balance the supply and demand and extract the largest profits.  Cost function based automated market maker with prices that sum to greater than one have been studied in \cite{Othman:2010,Abernethy:2010,Othman:2011} which implies these market makers that can turn a profit. These market makers do not, however, optimise the amount of profits they extract from traders. The Bayesian Market Maker of \cite{brahma2012bayesian} does not provide worst-case guarantees, nor this it optimise it's profits, but rather provides narrow spreads in equilibrium. The optimal policy for market makers in equities is studied by \cite{mildenstein2012optimal}. Using stochastic dynamic programming, \cite{huang2012optimal}  show that a threshold inventory control policy is optimal with respect to an exponential utility criterion and a mean-variance trade-off model.

Cost function based market makers for prediction markets are based on
sequentially shared proper scoring rules. These are myopically incentive
compatible: that is if traders do not consider the effects of their trades on
other players beliefs or on the market makers future actions then proper scoring
rules incentivise players to reveal their true beliefs. If traders can interact
multiple times with the market maker and act strategically, proper
scoring rules are not enough to incentivise traders reveal their true
beliefs \cite{chen2009gaming}. Incentive compatible bandit mechanisms are studied in \cite{gonen2007incentive}. The focus of this paper is where an (possibly adversarially arranged) sequence of traders get to interact only once with the market maker, which is sufficient to make the proper-scoring rule interpretation of the cost-function based market makers incentive compatible.

Using a bandit framework to attack a pricing problem goes back in the economics literature to \cite{rothschild1974two} who considers the case with discounting.  Special cases of the problem has more recently been analysed using modern methods by \citet{besbes2009dynamic}. A specific version of online posted-price auctions have been analysed  using adversarial bandits by  \citet{kleinberg2003value}.

\section{Framework}\label{sec:framework}

We now give a high level overview of our bandit market making framework. 
Details and assumptions about the various components are given in the 
subsections below.

We consider a setting with $n$ mutually exclusive and collectively exhaustive
outcomes, and $T$ time periods of trading. At each time period $t=1,\ldots,T$ 
the market maker has an obligation vector $q^t$ and must choose a cost function
$C^t$ from a predefined set $\Ccal$ which defines its pricing behaviour for
that round.
Traders sequentially interact with the market maker by buying portfolios of 
contracts at prices set by the market maker.
The aggregate purchases $s^t$ of the sequence of traders that 
arrives during a period $t$ shifts the market maker's obligation vector to 
$q^{t+1} = q^{t} + s^t$, its obligation vector for the beginning of round $t+1$. 

As described in the introduction, the explore vs.\ exploit trade off the market
maker faces in choosing a new cost function each round can be viewed as a
bandit problem.
To make use of algorithms from the bandit literature and their associated
guarantees, we first formalize our cost function based market and derive
some basic results that will be needed in Section~\ref{sec:bandits}.

\subsection{Cost functions}\label{sec:cost-functions}

The pricing behaviour of a market maker is defined through a continuous 
\emph{cost function}
$C:Q\to\RR_{+}$ where $Q$ is a closed, bounded subset of $\RR_+^n$. 
The cost function assigns a monetary value $C(q)$ to each \emph{market position}
or \emph{obligation} described by a vector $q\in Q$. 
The set $Q$ represents the positions a market maker using such a $C$ is willing 
to take prior to trading.
Each component $q_{i}$ is the total size of the market maker's obligation in 
case event $i$ occurs.

If the market is in position $q$ and a trader wants to
buy a portfolio of $s \in \RR_+^n$ shares, the price the trader must pay is 
$C(q+s)-C(q)$. 
This means the \emph{instantaneous price} per share for each security $i$ is
\emph{$\frac{\partial}{\partial q_{i}}C(q)$} and can be summarised by the
gradient vector 
\[	
	\pi(q) 
	:= \nabla C(q) 
	= \left(
		\frac{\partial C(q)}{\partial q_{1}},
		\ldots,
		\frac{\partial C(q)}{\partial q_{n}}\right).
\]
More details about the properties and methods for constructing cost functions
can be found in \cite{Abernethy:2010}.

It is important to note that our worst-case loss is bounded by the worst-case loss of the cost-function that provides the most liquidity in the set under consideration. The bandit algorithm guarantees we have vanishing regret relative to the optimal fixed choice of $C$ within our set.

\subsubsection{Market maker value}

A market maker attempts to balance the income it receives by selling contracts 
against the potential liabilities it faces when an outcome occurs and the 
corresponding contracts must be honoured.
Hence, we define the \emph{value} $V_q(C,i)$ of a market maker with cost-function $C$ and market position $q$ when outcome $i$ obtains to be
\[
	V_q(C,i) := C(q) - \inner{q}{e^{(i)}} = C(q) - q_i
\]
where $e^{(i)}$ is the $i$th basis vector in $\RR^n_+$ 
(\ie, $e^{(i)}_j = 0$ for $i\ne j$ and $e^{(i)}_i = 1$).
This quantity is the difference between the total income the market maker 
obtained by accepting trades to move to position $q$ and the cost of those 
obligations should outcome $i$ occur.
We abuse notation slightly and also define the 
value $V_q(C,p)$ where $p \in \Delta^n$ to be 
\[
	V_q(C,p) := \EE_{i\sim p}[ V_q(C, i) ] = C(q) - \inner{q}{p}
\] 
which is the difference between the total income of the market maker and the 
expected cost given a distribution $p$ over outcomes.

\subsubsection{Assumptions and properties}

We put some natural restrictions on the cost function, similar to those used
by \citet{Abernethy:2010} and \citet{Othman:2011}: 
\begin{itemize}
	\item[C1] \textbf{Convexity}:
		$C(\lambda q + (1-\lambda) q') \le \lambda C(q) + (1-\lambda)C(q')$ 
		for all $q,q'\in Q$ and all $\lambda \in [0,1]$.
	\item[C2] \textbf{Monotonicity}:
		$C(q) \ge C(q')$ for all $q,q'\in Q$ such that $q \ge q'$ (that is
		$q_i \ge q'_i$ for $i=1\ldots n$).
	\item[C3] \textbf{Bounded Loss}:
		$\sup_{q\in Q} \max_i q_i - C(q) < \infty$.
\end{itemize}

Since loss and value are complementary, we note that the bounded loss condition 
can be equivalently written as a lower bound on the worst-case value of a market
maker.
\begin{itemize}
	\item[C3'] \textbf{Bounded Value}:
		$\inf_q \min_i V_q(C, i) > -\infty$.
\end{itemize}

In addition to the above conditions, we also require that the cost function always offer prices that are
\emph{potentially profitable for the market-maker} and have 
\emph{bounded prices}:
\begin{itemize}
	\item[C4] \textbf{Potentially Profitable}:
		$\sum_{i=1}^n \pi_i(q) \ge 1$ for all $q \in Q$, 
		and there exists a $q \in Q$ such that $\sum_{i=1}^n \pi_i(q) > 1$ 
	\item[C5] \textbf{Bounded Prices}:
		There exists a $P < \infty$ so that for all $q \in Q$ the 
		prices $\pi(q) \in [0,P]^n$.
\end{itemize}

These last two conditions are crucial for the rest of our analysis.
The potentially profitable condition C4 is similar to that of \citet{Othman:2010} 
and requires that prices sum to greater than one for for some $q$ and never to 
less than one.
That is, for some state of the quantity vector, the cost function should offer
a set of prices that if a uniform vector of assets is purchased by traders it
will make guarantee a profit for the market maker.

The bounded prices assumption C5 implies that the cost functions we consider are
Lipschitz continuous.
\begin{lemma}\label{lem:lipschitz-C}
	If $C$ is continuously differentiable with bounded prices then 
	for all $q,q' \in Q \subset \RR_+^n$
	\[
		|C(q) - C(q')| \le P \sqrt{n} \|q - q'\|
	\]
\end{lemma}
\begin{proof}
	The bounded prices assumption implies $\pi(q) = \nabla C(q) \in [0,P]^n$
	and so $\|\nabla C(q)\| \le P \sqrt{n}$.
	By convexity of $C$ we have for all $q,q'$ that 
	$C(q) \ge C(q') + \inner{\nabla C(q')}{q-q'}$.
	Without loss of generality, assume $C(q') \ge C(q)$ so then
	$\inner{\nabla C(q')}{q'-q} \ge C(q') - C(q) \ge 0$ and by Cauchy-Schwarz
	we have $P \sqrt{n} \|q'-q\| \ge |C(q') - C(q)|$ as required.
\end{proof}


\subsubsection{Overround Cost Functions}
\label{sub:overround}

Families of potentially profitable cost functions can easily be constructed 
from a given cost function $C$ with fair prices, by charging a fixed 
multiple $a > 1$ of the prices for $C$.
We term these \emph{overround cost functions} for the \emph{base function} $C$.
When discussing such families we will always assume or show that the base function 
satifies conditions C1--C5.

For example, if $C(q) = b\log\sum_i\exp(q_i/b)$ is the well known logarithmic 
market scoring rule (LMSR) \cite{Hanson:2003} with fixed parameter $b$, we can 
construct the single parameter family $\Ccal = \{ a C : a \in [1,\infty) \}$.
In this case, the prices charged by some $C = a C \in \Ccal$ at market 
position $q$ are simply
\[
	\pi_i(q)  = a \frac{\exp(q_i /b)}{\sum_i \exp(q_i/b)}
\]
which is just $a$ times the prices $\pi^{LMSR}(q)$ given by the LMSR cost function 
with the same liquidity parameter $b$.

We will refer to this family of potentially profitable cost functions built
upon the LMSR as the \emph{overround LMSR} cost functions.
It is straightforward to verify that overround LMSR cost functions satisfy
conditions C1-C5. Any LMSR cost function satisfies C1-C3 
(see, for example, \cite{Abernethy:2010}) and multiplication by a constant
$a$ does not affect its convexity, monotonicity, or boundedness. 
Condition C4 holds because $\sum_i \pi_i(q) = a \sum_i \pi^{LMSR}_i(q) = a > 1$
for all $q\in Q$.
Similarly, the overround LMSR functions have bounded prices because 
$\pi^{LMSR}(q) \in [0,1]^n$ implies $\pi(q) \in [0,a]^n$ and so choosing
$P = a$ witnesses condition C5.

\subsubsection{Cost Function Distance}

When treating a set of cost functions as arms in a bandit game in
Section~\ref{sec:bandits}, we require a metric measure of ``closeness'' between 
cost functions. 
The most convenient measure for our purposes is the metric derived from the supremum
norm, which takes the maximal difference in costs assigned to any market position by 
two given cost functions.

Formally, we define the metric
\[
	d_\infty(C,C') := \sup_{q\in Q} |C(q) - C'(q)|
\]
and consider two cost functions to be close if $d_\infty(C,C')$ is small.
It is easy to establish that $d_\infty$ is indeed a (pseudo-)metric:
$d_\infty(C,C') = d_\infty(C',C) \ge 0$ with $d_\infty(C,C) = 0$ and the
triangle inequality carries over from the triangle inequality of $|x-x'|$
on the reals.

For overround cost functions this distance behaves naturally -- scaling with
the difference in overround. Specifically, if $aC$ and $a'C$ are two overround
cost functions for the base function $C$ then 
$d_\infty(aC,a'C) = |a-a'| \sup_{q\in Q} |C(q)|$.

\section{Bandit Market Making}
\label{sec:bandits}

Before showing how for profit market making can be viewed as a bandit problem
we first briefly outline bandit problems in general and point out some
salient results.
The remainder of this section then motivates a natural reward function that 
allows market making to be cast as a bandit problem. 
We then show how this particular choice of reward function satisfies the
conditions of a theorem of \citet{Kleinberg:2005a} for 
a fully adversarial continuous-armed bandit problem, thus allowing us to
apply this result to our market making problem.

\subsection{Bandit Problems}

A \emph{bandit problem}\footnote{%
	The reader wanting more details about bandit problems and an excellent 
	overview of the related literature is referred to \cite{Cesa-Bianchi:2006}.}
consists of a game played over $T$ rounds. 
In each round $t$ the player must choose an action $a^t$ (or ``arm'') from a 
predefined set of alternatives $\Acal$. 
After each choice the player is given a reward $r^t(a^t)$ and play continues to 
the next round.  
The player's aim is to accrue the largest total reward at the end of the $T$ rounds.
Crucially, the rewards for the unplayed actions are not revealed each round so
the player is forced to balance exploring unplayed actions (in case they have
larger rewards) and repeatedly playing those actions which have given high rewards
in the past.

When analysing algorithms that play bandit games, a central quantity is the 
\emph{regret} of a particular sequence of actions $a^{1:T} := a^1,\ldots,a^T$. 
This is the difference between the total reward $R(a^{1:T})$ the player received
and the best total reward that would have been received if a single arm were
played repeatedly. That is, the regret of the sequence $a^{1:T}$ is
\[
	Regret(a^{1:T}) 
	:= \sum_{t=1}^T r^t(a^t) - \max_{a\in\Acal} \sum_{t=1}^T r^t(a).
\]

Typical analyses of bandit problems derive sublinear (i.e., $o(T)$) 
bounds on the worst case regret for particular algorithms and under various 
assumptions about the power of the player's adversary when choosing the sequence of 
reward functions $r^1, \ldots, r^T$.
The structure of the action space $\Acal$ (e.g., finite, convex, etc.) and the 
set of reward functions the adversary can choose from also play an important role
in the type of results that are obtained.

For reasons described in the next subsection, we will restrict our attention 
to what are called
\emph{fully adversarial} bandit problems on \emph{continuous action spaces}.
The latter condition just means that the set of actions available to the player
is uncountable and typically some compact subset of $\RR^d$. 
The former term means that the choice of reward function at
each round can depend on the history of choices made by the player.

Because of the difficulty of this setting, there are relatively few results.
Those that work on action spaces that are a subset of $\RR^d$ tend to place
strong restrictions on the form of the reward functions, such as requiring
them to be linear or convex.

The one result we are aware of that applies to more general reward functions 
on continuous action spaces is by \citet{Kleinberg:2005a}. This only
requires a technical condition on the set of reward functions
that broadly says that they do not vary too wildly.
Formally, a function $f : S \to \RR$ is said to be \emph{uniformly locally
Lipschitz} with constant $L>0$, exponent $\alpha\in(0,1]$, and restriction $\delta>0$
whenever $f$ satisfies, for all $u,u'\in S$ with $|u-u'|\le\delta$, $|f(u)
- f(u')|\le L|u-u'|^\alpha$. Following Kleinberg, we denote this class of
functions $ulL(\alpha,L,\delta)$.
Given this definition, we now state Kleinberg's result:
\begin{theorem}[\cite{Kleinberg:2005a}]
	\label{thm:kleinberg}
	Let $S$ be a bounded subset of $\RR$ and 
	$\Gamma \subset ulL(\alpha,L,\delta)$ 
	be a set of uniformly locally Lipschitz reward functions $r : S \to [0,1]$
	Then there exists an algorithm CAB that achieves regret
	$O\left(T^{\frac{\alpha+1}{2\alpha+1}}\log^{\frac{\alpha}{2\alpha+1}} T\right)$
	against adaptive adversaries.
\end{theorem}

We now describe how to interpret market making in this setting and 
introduce a parameterised reward function for market making that satisfies 
the uniformly locally Lipschitz condtion.

\subsection{Market Making as a Bandit Problem}

When viewed as a bandit problem, the task of choosing a sequence of cost 
functions $C^1, \ldots, C^t \in \Ccal$ imposes a number of constraints on the
type of algorithm we can use to solve it.

One unavoidable difficulty is the inherent path dependence of the market making
problem. In any realistic market scenario, each choice of cost function 
would yield different amounts of trade. Given a fixed obligation vector
$q^t \in Q$, this means two different choices of $C^t$ will result in 
different vectors $q^{t+1}$ in the subsequent round.
This means that any sequence of reward functions that takes into account 
both the choice of cost function and the obligation vector must necessarily 
depend on the history of cost function choices.
Thus, any interpretation of market making as bandit problem must assume
fully adversarial (i.e., non-independent) choices of rewards.

The second main difficulty is due to the richness of actions that the market
maker can take.
In its most general form, $\Ccal$ could be the set of all 
cost functions for some fixed domain $Q$ that meet conditions C1--C5.
However, this is too rich a class of actions to be handled by any existing
adaptive bandit results that we are aware of and restricting attention to
finite $\Ccal$ seems too restrictive, as the natural scale of the changes in the overround that should be searched over is a priori unknown.

To the best of our knowledge, the largest class of actions for which
adaptive regret guarantees are available for non-linear and non-convex rewards
are for those in Theorem~\ref{thm:kleinberg}. That is, the action space must
be (isomorphic to) a bounded subset of $\RR$.
A natural class of cost functions $\Ccal$ that map to such an action space are
the overround cost functions described in Section~\ref{sub:overround}.
In particular, we will restrict our attention to overround cost functions 
with bounded overround parameter. 
That is, $\Ccal = \{ a C : a \in [1,A] \}$ for some $A < \infty$ and choice of
base cost function $C$.
By identifying each cost function $aC \in \Ccal$ with $a \in [1,A]$ we meet the
requirement of Theorem~\ref{thm:kleinberg} that the action space be a 
bounded subset of $\RR$.

\subsection{Reward Function}

The final piece in our interpretation is the class of reward functions 
for the bandit problem.
We do so by first defining a state- and trade-dependent reward function
$r_q(C;s)$ for cost function $C$, state $q$, and trades $s$. 
We show that this function has a number of key properties and then use it
to derive the class 
$\Rcal = \{ r_q(C;s) : q \in Q, s \in \RR_+^n, q+s \in Q \}$.
The properties of $r_q(C;s)$ are then used to show that $\Rcal$ is
uniformly locally Lipschitz with $\alpha=1$ and thus meets the condtions
of Theorem~\ref{thm:kleinberg}.

The state- and trade-dependent reward we proposed is simply 
the difference between the 
worst-case expected values of the market maker before and after trading.  Worst case profits are the natural distribution free measure of profits. The change in their value in a given period is thus a natural way to reward the bandit for the trades that occur during that period.


Formally, the reward for using cost function $C$ in state $q$ for trades $s$ 
is
\begin{equation}\label{eq:reward}
	r_q(C; s) := 
		\min_{p\in\Delta^n} V_{q+s}(C, p) 
		- \min_{p'\in\Delta^n} V_q(C, p').
\end{equation}

We now show that this reward functions behaviour is reasonable.

\begin{theorem}
	The reward function $r_q(C;s)$ defined in (\ref{eq:reward})
	has the following properties:
\begin{enumerate}
	\item[R0] \textbf{Zero Calibrated}:
		For all $C$, $q$ the reward function satisfies
		\(
			r_q(C; 0) = 0.
		\)
	\item[R1] \textbf{Path Independent}: 
		For all $C$, $q$, $s_1$, and $s_2$ the reward function must satisfy
		\[
			r_q(C; s_1 + s_2) = r_{q+s_1}(C; s_2) + r_q(C; s_1).
		\]
	\item[R2] \textbf{Overround Compatible}:
		For all $q = k.\ones$ and $s = k'.\ones$ with $k,k' \ge 0$,
		and homogenous cost-function $C$ with overround 
		$\alpha$ (\ie, $C(k.\ones) = k.C(\ones) = k.(1+\alpha)$) the reward 
		satisfies
		\[
			r_q(C; s) = k'.\alpha
		\]
\end{enumerate}
\end{theorem}

\begin{proof}
	R0 holds trivially since when $s=0$ the two minimisations in (\ref{eq:reward})
	are identical and thus cancel leaving $r_q(C; 0) = 0$.

	R1 is shown by observing that 
	\begin{align*}
		r_{q+s_1}(C;s_2) + r_q(C; s_1)
		=& \min_p V_{q+s_1+s_2}(C,p) - \min_{p'} V_{q+s_1}(C,p') \\
		 & + \min_{p''} V_{q+s_1}(C,p'') - \min_{p'''} V_{q}(C,p''') \\
		=& \min_p V_{q+s_1+s_2}(C,p) - \min_{p'''} V_q(C,p''') \\
		=& r_q(C; s_1 + s_2)
	\end{align*}
	since the middle two terms are identical optimisations and hence cancel.

	To establish R2 we first observe that the value of a cost-function $C$ 
	and position $q$ is independent of $p$ when $q = k.\ones$ for some
	$k \ge 0$. 
	This is because 
	$V_q(C,p) = C(q) - \inner{q}{p} = C(q) - k.\inner{\ones}{p} = C(q) - k$.
	If $C$ is homogeneous then for such $q$ we have $C(q) = C(k.\ones) = k.C(\ones)$
	and so $V_q(C,p) = k.(C(\ones) - 1)$ for all $p$.
	So for such $C$, $q$ and for $s = k'.\ones$ we have
	\begin{align*}
		r_q(C;s) 
		=& \min_p V_{q+s}(C,p) - \min_{p'} V_q(C,p') \\
		=& (k+k').(C(\ones)-1) - k.(C(\ones) - 1) \\
		=& k'.(C(\ones) - 1) 
	\end{align*}
	where the second term in the last line is just the overround $\alpha$ for $C$.
\end{proof}

R0-R2 were derived from considering how a reward function should
relate to the cost-function in cases when the uncertainty over outcomes is 
irrelevant or not present.
R0 guarantees that if a choice of cost-function results in no trading then
no reward is received.
R1 ensures that the order of trades for a fixed cost-function does not affect
the overall reward. Much like path independence for cost-functions themselves.
R2 comes from the observation that if the market maker is in a position with a 
value that is outcome independent (\ie, $q = k.\ones)$ and a trader buys an outcome
independent bundle of shares $s$ then the reward should be exactly the overround
for $C$ on $s$ (\ie, $k'.\alpha$). 

The next lemma gives a bound on how much $r_q(C;s)$ can vary as a function of
a perturbation in $C$ and a change in $s$. This is need in 
Theorem~\ref{thm:lipschitz} below to establish the smoothness of the reward
class $\Rcal$.

\begin{lemma}
	For each choice of base function $C_0$ with prices bounded by $P$
	and $A>1$ let $\Ccal =  \{ a C_0 : a \in [1,A] \}$.
	Then the function $r_q(C; s)$ satisfies
	\[
		\left| r_q(C; s) - r_q(C'; s') \right|
		\le 
		2 d_\infty(C,C') + (A.P.\sqrt{n} + 1)\|s - s'\|
	\]
	for $C,C' \in \Ccal$ and for all
	$q \in Q$ and $s,s' \in \RR_+^n$ such that $q+s \in Q$ and $q+s'\in Q$.
\end{lemma}

\begin{proof}
	Let $\bq := q + s$ and $\bq' := q + s'$ and consider the 
	difference
	\begin{align*}
		|r_q(C;s) - r_q(C';s')|
		  =& | C(\bq) - C'(\bq') + [ C'(q) - C(q) ] 
		     + [ \min_{p'} \inner{p'}{\bq'} 
		     - \min_{p} \inner{p}{\bq} ] \\
		\le& \underbrace{| C(\bq) - C'(\bq') |}_{T_1}
		     + \underbrace{| C'(q) - C(q) |}_{T_2} 
		     + \underbrace{|\min_{p'}\inner{p'}{\bq'}-\min_{p}\inner{p}{\bq}|}_{T_3}. 
	\end{align*}

We now argue that each of the three terms are bounded by some multiple of 
$d_\infty(C,C') := \sup_{q\in Q} |C(q) - C'(q)|$.
Clearly, $T_2 \le d_\infty(C,C')$ by definition of $d_\infty$.
For $T_1$ we see
\begin{align*}
	T_1
	=& |C(\bq) - C(\bq') + C(\bq') - C'(\bq')| \\
  \le& |C(\bq) - C(\bq')| + |C(\bq') - C'(\bq')| \\
  \le& A.P.\sqrt{n} \|s - s'\| + d_\infty(C,C').
\end{align*}
The last term is by the definition of $d_\infty$,
while the first term is because $\bq - \bq' = s - s'$ and 
Lemma~\ref{lem:lipschitz-C}, since every $C \in \Ccal$ has prices bounded
by $A.P$.

Finally, for $T_3$ suppose without loss of generality that 
$\min_{p'} \inner{p'}{\bq'} \ge \min_{p} \inner{p}{\bq}$ and that $p^*$
satisfies $\inner{p^*}{\bq} = \min_{p} \inner{p}{\bq}$.
Then 
\(
	T_3 
	= \min_{p'} \inner{p'}{\bq'} - \inner{p^*}{\bq} 
  \le \inner{p^*}{\bq'} - \inner{p^*}{\bq} 
	= \inner{p^*}{s'-s} 
\).
Thus, by Cauchy-Schwarz, we have $T_3 \le \|p^*\|\|s-s'\|$ and because $\|p\| \le 1$ for all $p\in\Delta^n$ this gives $T_3 \le \|s-s'\|$.
Putting the above bounds together gives the required result.
\end{proof}

\subsection{The continuous case}

To define a class of reward functions in terms of $r_q(C;s)$ we let
$\Rcal = \{ r_q(C;s) : q \in Q, s \in \RR_+^n, q+s \in Q \}$.
Each $r \in \Rcal$ is a reward function for cost functions with an implicit
state vector $q$ and trade vector $s$.
The interpretation of this class in a bandit context is that reward
functions are drawn semi-adversarily in that in any state $q$,
an adversary is free to choose the trade vector $s$, resulting in the
selection of the reward function $r_q(C;s) \in \Rcal$.
However, in the subsequent round, the adversary will then be forced to choose 
a reward function of the form $r_{q+s}(C;s')$.
 
If we are to assume that the trades depend on the choice of cost function then
we need to be able to guarantee that this difference does not change too
rapidly if the choice of cost function only changes a small amount 
(as measured by $d_\infty$).
To achieve this, we further constrain the adversary assume that during any 
round of trade
the largest possible difference in trades the market maker can face 
if choosing between using $C$ or $C'$ is bounded by $K.d_\infty(C,C')$
for some constant $K$.
We will call this the \emph{$d_\infty$-smooth trades} assumption and
use it in the statement of our main result.


With the above assumptions in place, the following result states that it is 
possible to for a bandit market maker to achieve vanishing regret relative
to a single choice of overround cost function, given the previously
discussed regularity assumptions about the variability of trades with respect
to choice of cost function by the market maker.
This means, despite not knowing what the distribution of traders' demands are
it is possible for the bandit market maker to asymptotically extract a 
comparable amount of expected profit as if the best cost function in the class for 
traders with those demands was known in advance.

\begin{theorem}\label{thm:main}
	Let $A>1$, $C_0$ be a fixed base cost function, and let 
	$\Ccal_{\text{over}} = \{ a\,C_0 : a \in [1,A] \}$
	be the family of overround cost functions with metric 
	$d_\infty(C,C') := \sup_{q\in Q} |C(q) - C'(q)|$.
	Then there exists a bandit market maker (using the CAB algorithm) that 
	achieves regret $O(T^{2/3} \log^{1/3} T)$ against 
	$d_\infty$-smooth trades.
\end{theorem}

\begin{proof}
The proof of this theorem hinges on showing that the reward functions
$\Rcal$ can be transformed into rewards for a bandit game that are
$[0,1]$-valued and uniformly locally Lipschitz with $\alpha =1$.
Once this is done, the algorithm CAB can be used to choose arms $a\in[1,A]$
corresponding to cost functions $C = a\,C \in \Ccal_{\text{over}}$. 
The regret bound for the game over $[1,A]$ then carries over to the game on
$\Ccal_{\text{over}}$.

To see that the reward functions in $\Rcal$ can be mapped to functions with 
range $[0,1]$, note that $C_0$ is a continuous function over a closed,
bounded set $Q$ and thus must be bounded by some $B$. 
Therefore and $C = a C_0$ must also be bounded by $AB$ since $a < A$. 
Thus, 
\[
	r_q(C;s)
	= \min_p V_{q+s}(C,p) - \min_{p'} V_q(C,p') 
	\le AB - 1 + K
\]
where $-K$ is the lower bound guaranteed by condition C3' for the cost 
function $C_0$.
A lower bound is similarly established and so every $r_q(C;s)$ takes
on values in some bounded range which can be mapped to $[0,1]$.

The uniformly local Lipschitz property for rewards from $\Rcal$ holds
due to Lemma~\ref{lem:lipschitz-C} and the $d_\infty$-smooth trades 
assumption. Specifically, if $r = r_q(\cdot;s)$ and $r' = r_q(\cdot;s')$
are two reward function from $\Rcal$ then Lemma~\ref{lem:lipschitz-C}
guarantees that
\[
	|r(C) - r'(C')|
	\le 2 d_\infty(C,C') + (A.P.\sqrt{n} + 1)\|s-s'\|
	\le [2 + S.(A.P.\sqrt{n} + 1)] d_\infty(C,C')
\]
since $\|s-s'\| \le S.d_\infty(C,'C)$ by the smoothness assumption.
Thus, the class $\Rcal$ is uniformly locally Lipschitz with $\alpha = 1$,
as required.
\end{proof}


The above result shows that the total reward of a market maker playing the 
bandit strategy is not much worse ($O(T^{2/3} \log T^{1/3})$) than the 
total reward for the best single choice of cost function in hindsight.
Since the reward at each round measures the worst-case  



\section{Simulations}

A discrete version of a bandit market maker was implemented, using a overround LMSR as the cost function and EXP3 as the bandit algorithm. We explore the behaviour of the cumulative guaranteed profits of this market maker in simulations to asses it's performance. All simulations are done by generating the same sequence of traders with beliefs and budgets, they sequentially interact with the market makers cost function once and purchase optimal bundles given their budget and beliefs. The liquidity parameter $b=10$. Each trader had a unit budget, and one trader arrived on each time period. A range of  values for the overround was included, ranging from 1.05 to 1.8 by doubling the interval between each. Beyond the bandit market maker, each of the market makers in the comparator class was evaluated using the same sequence of traders. All market makers had a uniform starting asset vector set at 0, and there where two possible outcomes.

\subsection{IID trader beliefs}

To evaluate the baseline behaviour of the market maker we simulate traders beliefs being drawn from a uniform distribution in the range $[0,1]$ for 400 periods. Note that the maximal possible profitability without knowledge of the individual traders in the sequence demands is to set prices of 0.75 for both of the assets (effectively an overround of 1.5) and not move them in response to trades. After 400 traders interact wight he market, half of them will have purchased shares, so maximum profits are 100. It is interesting to note that despite an overround of 1.5 not being considered, and the liquidity being relatively limited (as opposed to effectively infinite in the optimal case) the best overround market maker is able to obtain 2/3 of the maximal profits, while the bandit market maker using EXP3 obtains 2/5 of the maximum profits. 

\includegraphics{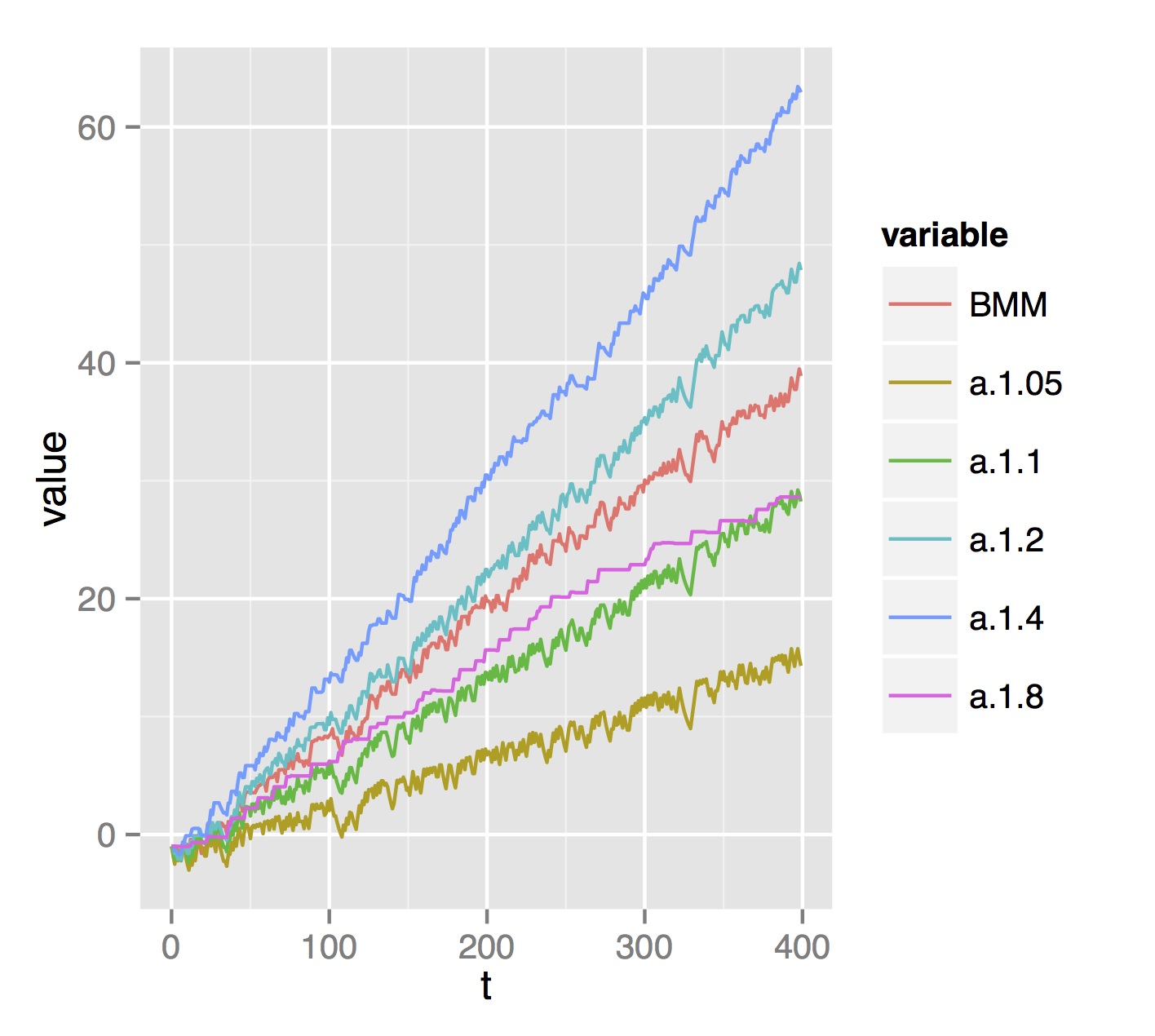}

\subsection{Adapting to a shock}

To evaluate the performance of the market maker to a shock we again start the market makers with a uniform price vector, but traders beliefs are initial distributed uniformly in the range $[0.15,0.35]$ for the first 500 periods, after that they switch to a regime in which they are uniformly drawn from the interval $[0.65,0.85]$. The initial adjustment of prices downwards towards the 0.25 equilibrium during the first half is accompanied by a large potential worst case loss. An adjustment in the opposite direction when the regime switches in the midway point leads to a sequence of time periods during which all market makers meet their upper bound of per period profits, since the prices are very attractive to arriving traders who spend their entire budget, while the worst case loss of the market maker remains unaffected as they are purchasing the asset that was not purchased int he first half of the trading period. 
\includegraphics{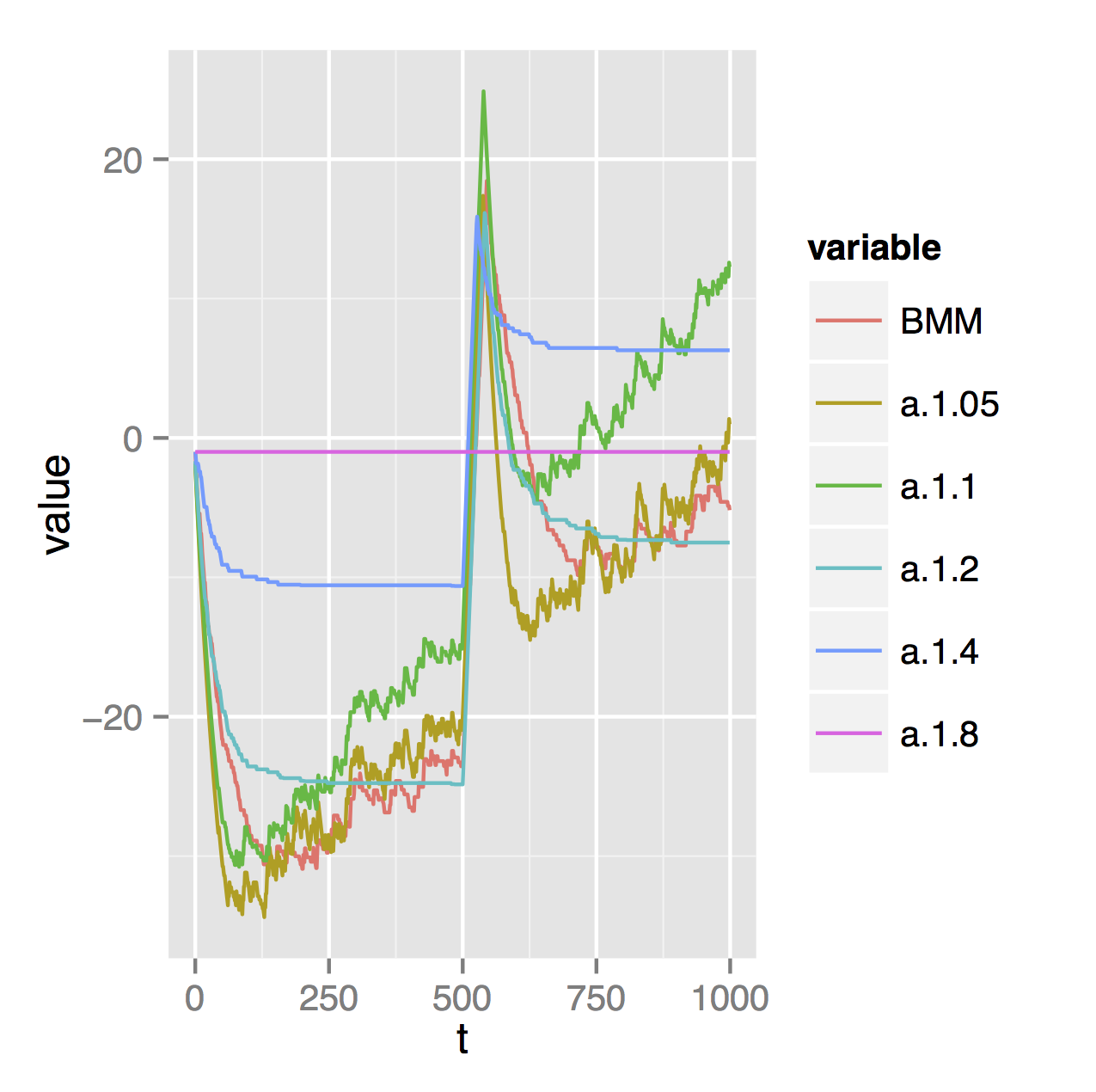}

\section{Conclusion and future work}

Our approach preserves the model-free worst case guarantees of the prediction markets literature, while simultaneously being able to provide guarantees of profit optimality, these profit optimal guarantees are, however, relative to a particular class of market makers. It would be interesting to understand in future work how far this class of market makers can be from optimal within the set of market makers that have bounded risk. The range of bandit algorithms in the literature is ever growing, and further work in exploring which are best suited for market making environments would also be of substantial interest. Evaluation of market makers within this framework in real life for-profit market making would naturally also be of interest.



\bibliography{../markets}


\end{document}